# Astro2010 Science White Paper (SSE)
## *Stellar-Mass Black Holes and Their Progenitors*
J. Miller, Uttley, Nandra, Barret, Matt, Paerels, Mendez, Diaz-Trigo, Cappi, Kitamoto, Nowak, Wilms, Rothschild, Smith, Weisskopf, Terashima, Ueda

If a black hole has a low spin value, it must double its mass to reach a high spin parameter (Volonteri et al. 2005). Although this is easily accomplished through mergers or accretion in the case of supermassive black holes in galactic centers, it is impossible for stellar-mass black holes in X-ray binaries. Thus, the spin distribution of stellar-mass black holes is almost pristine, largely reflective of the angular momentum imparted at the time of their creation. This fact can help provide insights into fundamental questions:

- **What is the nature of the central engine in supernovae and gamma-ray bursts?**
- **What was the spin distribution of the first black holes in the universe?**

Optical depth effects make it difficult to directly probe the central engines in supernovae (SNe) and gamma-ray bursts (GRBs). Especially in the case of GRBs, we only see the surface of last scattering in a highly relativistic jet, far from the black hole that is created through the collapse of a massive star, the merger of two neutron stars, or the merger of a neutron star and a black hole. However, the angular momentum of the black hole remnant is an important clue that can greatly aid in efforts to reverse-engineer the most powerful explosions in the universe (Heger et al. 2002, Nomoto et al. 2003, Gammie et al. 2004, Woosley & Heger 2006). **Do all explosions put equal energy into black hole spin? Does spin power the jets in GRBs?**

Similarly, understanding the spin of natal black holes in the early universe is extremely important. Spinning black holes convert accretion energy into radiation more efficiently, for instance, and would have supplied young host galaxies with more ionizing radiation that could impact star formation. While it is presently unclear whether a single massive black hole formed in early galaxies, or if a supermassive black hole was built out of the merger of many intermediate-mass black holes resulting from Population III stars, *stellar-mass black holes represent our best opportunity for understanding the distribution of spins that result from a single collapse event.*

It is possible that many factors play into determining how much angular momentum is imparted to a black hole formed in a SNe or GRB. However, fundamental properties such as the mass and and the angular momentum of the progenitor star must be important (Heger et al. 2002, Woosley & Heger 2006). The most sophisticated treatment of this problem suggests that single collapse events should create black hole with spins of $a = 0.75$-$0.90$ (Gammie et al. 2004), where $a = cJ/GM^2$. Presently, only a limited number of observations have obtained data that can be used to constrain the spin of stellar-mass black holes. The most complete study to-date only includes eight black holes (Miller et al. 2009); Figure 1 plots the resulting distribution, and includes two additional results. This modest histogram is the result of using relativistic disk lines and disk reflection spectra to constrain black hole spin. Clearly, to make a strong test of the theoretical prediction above, to provide a strong observational grounding for the nature of SNe/GRB central engines, and to understand the first black holes, the number of spin measurements must increase by an order of magnitude.

*IXO represents the best means of obtaining the greatest number of new black hole spins because it offers 5 independent means of measuring spin: relativistic disk lines, the disk continuum, high frequency quasi-periodic oscillations, polarimetry, and studies of X-ray reverberation.*

**Relativistic Disk Lines**

Irradiation of the accretion disk by hard X-rays produces emission lines and a characteristic disk reflection spectrum (for a review, see Miller 2007). The most prominent line is typically Fe K, due to the abundance and fluorescence yield of iron. The disk reflection spectrum is typified by an apparent flux excess peaking between 20-30 keV, which is actually due to Compton back-scattering. Relativistic Doppler shifts and gravitational red shifts endemic to the inner disk around black holes act to skew the shape of disk lines and the reflection spectrum. The shifts grow more extreme with increasing black hole spin, as the innermost stable circular orbit (ISCO) extends closer to the black hole. The clear imprints of special and general relativity on the line profile are thus used to measure black hole spin. Because the line shifts scale with gravitational radii ($GM/c^2$), the mass of a given black hole and its distance are not needed to measure its spin using this technique.

Relativistic disk lines are common in Seyfert AGN (e.g. Brenneman & Reynolds 2006, Miniutti et al. 2007), in stellar-mass black holes (Miller 2007; Miller et al. 2009), and even in the spectra of accreting neutron stars (Bhattacharyya & Strohmayer 2007, Cackett et al. 2008). A separate white paper describes how IXO spectroscopy can measure spin in 300 AGN and thereby constrain galaxy merger and accretion histories – important aspects of the coevolution of black holes and host galaxies. ***The commonality of disk lines permits comparisons of the relativistic regime around compact objects across a factor of $10^6$ in mass.***

IXO is ideally suited to measuring spin in stellar-mass black holes using relativistic disk lines and disk reflection, owing to the spectral resolution of the X-ray microcalorimeter spectrometer (XMS) and the flux and timing capabilities of the high time resolution spectrometer (HTRS). Figure 2 illustrates the relationship between spin, the ISCO, and relativistic disk lines.

**The Accretion Disk Continuum**

Thermal continuum emission from the accretion disk may be used to measure the spin of stellar-mass black holes. An accretion disk around a spinning black hole is expected to be hotter and more luminous than a disk around a hole with low spin, because the ISCO is deeper within the gravitational potential (see Figure 3). New spectral models have recently been developed that exploit these changes in the shape of the continuum to measure spin (Davis et al. 2005, Li et al. 2005). If the mass and distance to a black hole are known, these models may be applied to spectra in order to measure the spin of a stellar-mass black hole (see, e.g., McClintock et al. 2006, Shafee et al. 2006). By virtue of their sensitivity and spectral resolution, the XMS and HTRS aboard IXO are both well suited to measuring black hole spins using the disk continuum.

**X-ray Quasi-Periodic Oscillations**

The X-ray flux observed from black hole X-ray binaries is sometimes modulated at a frequency commensurate with Keplerian motion at the ISCO. The oscillations are not pure, but rather have a small width due to small variations in frequency and phase jumps – as expected for gas orbiting in a real fluid disk with internal viscosity. Observed frequencies follow a 1/M scaling (Remillard & McClintock 2006; see Figure 4), strongly suggesting a relativistic origin. Indeed, frequencies are sometimes observed in a 3:2 frequency ratio, consistent with parametric resonances in GR (Abramowicz & Kluzniak).

A few models are able to account for the frequencies observed; all of them require GR and each points to a high spin value when QPOs are observed in a 3:2 frequency ratio (see Figure 4). The large collecting area of IXO, combined with the energy range and time resolution of the HTRS, make IXO an apt next-generation X-ray timing mission. New detections of X-ray QPOs in black holes will yield new spin measurements and will elucidate the details of the modulations.

**X-ray Polarimetry**

Thermal X-ray emission from the accretion disc is likely to be significantly polarized. In the case of Newtonian gravitation, symmetry considerations demand that disk emission can only be parallel or perpendicular to the disk. Strong gravity effects distinctively modify the polarization properties, causing the polarization angle to appear rotated to the distant observer. The rotation is larger at smaller radii, where the disc temperature is higher, giving a unique energy dependence of the polarization angle (Stark & Connors 1977; Dovciak et al. 2008; Li et al. 2009). The degree of this dependence is related to the spin of the black hole. The plots in Figure 5 illustrate how IXO can measure black hole spin based on the polarization of thermal emission from the disk.

**X-ray Reverberation**

Measuring the light travel time between flux variations in the hard X-ray continuum and the lines that it excites in the accretion disk provides a model-independent way to measure black hole spin. The time delay simply translates into distance for a given geometry. If a black hole has a low spin parameter, iron emission lines in the 6.4-6.97 keV range should have a characteristic lag of approximately 6 $GM/c^3$; if the black hole is rapidly spinning, lags can be as short as 1 $GM/c^3$.

Close to spinning black holes, the path light takes will be strongly impacted by spacetime curvature. When very close to the black hole, an otherwise isotropic source of hard X-ray emission will have its flux bent downward onto the disk. An observable consequence of these light-bending effects is a particular non-linear relationship between hard X-ray emission and iron emission lines (Miniutti & Fabian 2004). At present, there is tantalizing evidence for this effect in some Seyfert AGN (e.g. Miniutti, & Fabian 2004, Ponti et al. 2006) and stellar-mass black holes (e.g. Rossi et al. 2005).

The IXO HTRS has the time resolution, broad energy range, energy resolution, and flux tolerance needed to make careful studies of lags that can lead to spin measurements and clear detections of gravitational light bending. A study of the lags that can be detected with the HTRS is shown in Figure 6. The extraordinary sensitivity of this instrument will enable errors of less than 1 $GM/c^3$ for fluxes of approximately 1 Crab.

**Extending Our Reach with IXO**

The distribution of stellar-mass black hole spins represents a rare and vital window on the central engines in SNe and GRBs and the first black holes in the universe. Is it the degree of angular momentum imparted to the black hole that separates SNe and GRBs? How much ionizing flux were the first black holes able to supply to young host galaxies? Current X-ray observatories are beginning to measure spin parameters in a small number of sources. The variety of techniques open to observers with IXO provides the best means of obtaining the greatest possible number of spin measurements. Moreover, the sensitivity of IXO will make it possible to obtain spin measurements in faint Galactic sources that are beyond the reach of current missions, and make it possible to obtain spin measurements in bright stellar-mass black holes in nearby galaxies. IXO will increase the number of current spin measurements by an order of magnitude, revealing the nature of the central engine in GRBs and SNe, and the nature of the first black holes to inhabit young galaxies at high redshift.

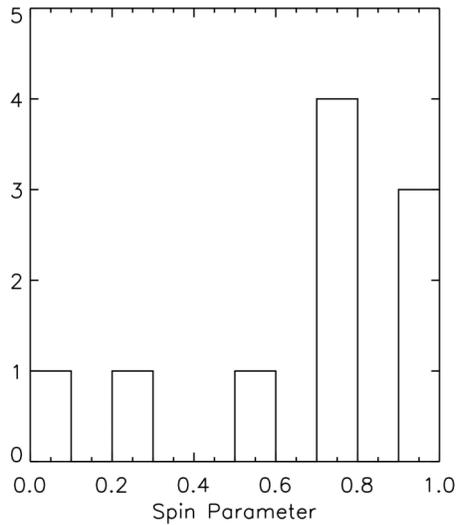

**Figure 1**: The histogram above shows a distribution of 10 black hole spins, measured using relativistic iron emission lines from the accretion disk (Miller et al. 2009, Reis et al. 2009, Blum et al. 2009). IXO will increase the number of stellar-mass black hole spin measurements by an order of magnitude, providing a unique window on the central engine in SNe and GRBs.

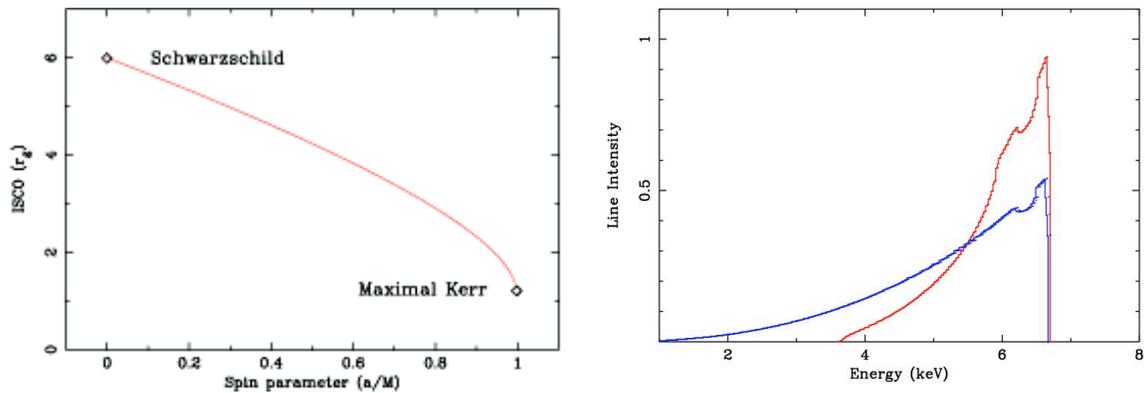

**Figure 2**: *Left*: Black hole spin can be measured by determining the innermost stable circular orbit of an orbiting accretion disk. All methods of determining spin that utilize the disk rely on this relation shown here. *Right*: the line profile in red is expected around a zero-spin Schwarzschild black hole; the more skewed line in blue is expected around a maximal-spin Kerr black hole. IXO will be able to reveal relativistic imprints on lines in sharp detail.

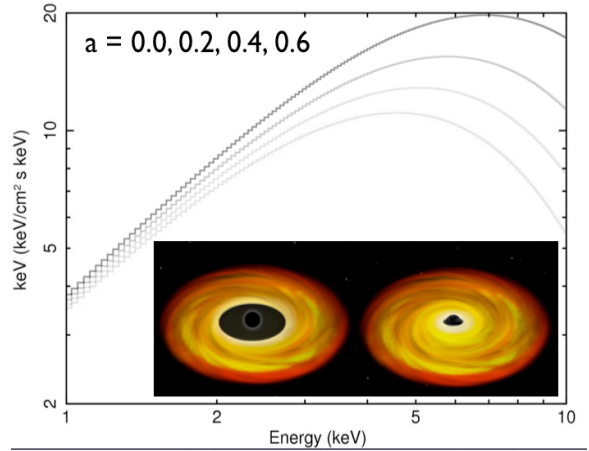

**Figure 3:** With increasing spin, thermal continuum emission from an accretion disk will become hotter and more luminous, because the ISCO moves closer to the black hole. The IXO XMS and HTRS are extremely well suited to measuring stellar-mass black hole spins using disk continua.

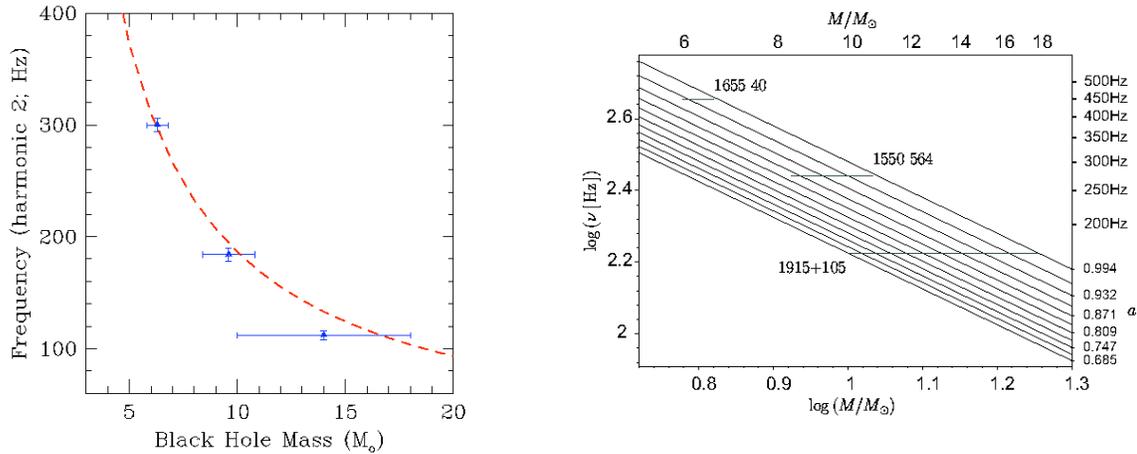

**Figure 4:** *Left*: High frequency QPOs in black hole X-ray binaries are consistent with a 1/M scaling, indicating a relativistic origin (Remillard & McClintock 2006). *Right*: The plot shown depicts QPO frequency (the upper frequency in a 3:2 resonance ratio) versus black hole mass. Parametric resonance models for the QPOs seen in a 3:2 frequency ratio are plotted over the data (Abramowicz & Kluzniak 2004). The IXO HTRS is designed to detect high frequency signals like these.

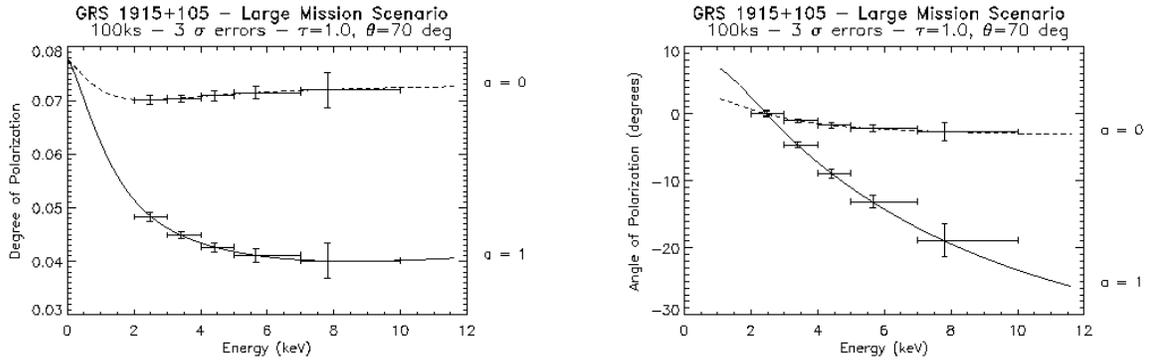

**Figure 5**: The plots above illustrate how the degree of polarization and angle of polarization differ for disks around a non-spinning (a=0) and maximally spinning (a = 1) black hole (Dovciak et al. 2008). IXO will fly a sensitive X-ray polarimeter that can use these signatures to measure spin in stellar-mass black holes. The plots above depict the results expected from a 100 ksec observation.

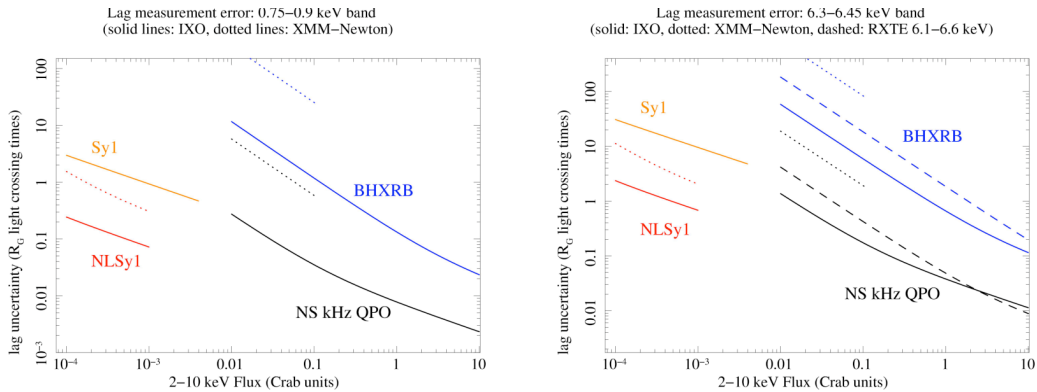

Figure 6: The IXO HTRS will be able to detect lags commensurate with the very shortest length scales around compact objects of all masses. This ability will provide another means of measuring black hole spin parameters. The left panel shows figures of merit for the Fe L range, and the right panel shows figures of merit for the Fe K band. In each plot, the uncertainty in the lag between the continuum and band of interest (in units of $GM/c^3$) is plotted versus source flux, for a variety of source types.